\documentclass[draft]{agujournal2019}
\usepackage{url} 
\usepackage{lineno}
\usepackage[inline]{trackchanges} 
\usepackage{soul}
\draftfalse

\journalname{Geophysical Research Letters}

\begin{document}

%
%

\title{Barchan dunes cruising dune-size obstacles}

\textcolor{blue}{An edited version of this paper was published by AGU.\\
	Assis, W.R., Borges, D.S., Franklin, E.M., Barchan dunes cruising dune-size obstacles. Geophysical Research Letters, 50, e2023GL104125, 2023. Published under the CC BY 4.0 license, DOI 10.1029/2023GL104125.\\
	To view open-access paper, go to https://doi.org/10.1029/2023GL104125.}

%
%




\authors{W. R. Assis\affil{1}, D. S. Borges\affil{1}, E. M. Franklin\affil{1}}


\affiliation{1}{School of Mechanical Engineering, UNICAMP - University of Campinas,\\
Rua Mendeleyev, 200, Campinas, SP, Brazil}




\correspondingauthor{Erick M. Franklin}{erick.franklin@unicamp.br}




\begin{keypoints}
\item We show that subaqueous barchans can be blocked, bypass, or pass over  dune-size obstacles
\item In some cases, barchans can split in two or more bedforms
\item We propose that the barchan behavior depends basically on two dimensionless parameters
\end{keypoints}

%
%

%
%


\begin{abstract}
We investigate the behavior of subaqueous barchans reaching dune-size obstacles by carrying out experiments where we varied the obstacle shape and size, the flow strength, and the grains' properties. We found that a subaqueous barchan can pass over or bypass a dune-size obstacle, or even be blocked, with some intermediate situations. In the bypass cases, the original barchan can split in two or more bedforms, redistributing sand in space. Finally, we propose a classification map in which the barchan behavior depends basically on two dimensionless parameters. Our results represent a step toward understanding how barchans behave in the presence of large obstacles, such as retaining walls, tubes and bridge pillars.
\end{abstract}

\section*{Plain Language Summary}
This paper is devoted to crescent-shaped dunes, known as barchans, that are found on Earth, Mars and other celestial bodies, with roughly the same morphology but different scales. In some cases, barchans can approach an obstacle of comparable size, such as houses and buildings in the aeolian case, and bridge pillars and submarine structures in the aquatic case. In order to investigate that, we carried out experiments in a water channel where granular heaps developed into barchans that approached obstacles of different shape and size. We found that barchans can be blocked, bypass or pass over dune-size obstacles, and that bypassing barchans can split in two or more bedforms, which means a significant redistribution of sand in space. Based on the experiments, we propose a classification map in which the barchan behavior depends on two dimensionless parameters. Our results shed light on how barchans behave in the presence of large obstacles, helping us to predict the outcomes of dunes interacting with objects in other environments, and design safer and eco-friendly structures.

\section{Introduction}

Barchans are crescent-shaped dunes that grow in gaseous or liquid environments under one-directional flows and limited amount of available grains, being frequently found on Earth, Mars and other celestial bodies \cite{Bagnold_1, Herrmann_Sauermann, Hersen_3, Elbelrhiti, Claudin_Andreotti, Parteli2, Courrech}. In common, barchans found in different environments share roughly the same morphodynamics, though variations associated with hilly terrains \cite{Finkel, Bourke, Parteli4}, wind changes \cite{Finkel, Bourke, Parteli4}, barchan-barchan collisions \cite{Long, Hersen_5, Bourke, Vermeesch, Parteli4, Assis}, or polidispersity \cite{Alvarez6, Assis3} have been observed. Length and time scales of barchans, however, are very different, going from centimeters and minutes under water to kilometers and millenniums in the Martian atmosphere \cite{Hersen_1, Claudin_Andreotti}. 

Although barchans are often organized in long fields, they can in some cases approach an obstacle of comparable size, such as houses and buildings in the aeolian case, and bridge pillars and submarine structures in the aquatic case. In addition to affecting human activities, this situation represents a threat to biodiversity \cite<such as in Len\c{c}\'ois Maranhenses, Brazil, where barchans and barchanoids reach grasslands and mangroves, and are close to villages,>{Amaral}, and a challenge for predicting the outcomes of long-time migrating dunes \cite<such as barchans and barchanoids interacting with large obstacles on Mars,>{Breed, Urso, Roback}. Besides, it can be an additional source of variations in the morphology of barchans. However, very few works investigated the effect of large obstacles on the behavior of sand dunes, and none of them specifically for barchans.

To the authors' knowledge, the only experimental investigation of dunes interacting with large obstacles is the one reported in \citeA{Bacik} for two-dimensional (2D) dunes. In that work, the authors carried out experiments in a narrow Couette-type circular channel, where different 2D obstacles were placed on the bottom and obstructed the dune path. They found that dunes either cross over the obstacle or remain trapped, and propose that the size and shape of the obstacle control the dune behavior via the flow structure near the obstacle. Their results are important for understanding when a 2D (or near 2D) dune can cross over a given obstacle, but the question for their three-dimensional (3D) counterparts remains open. For instance, is it possible for a bedform to circumvent a large obstacle, such as free-surface water flows do when flowing in subcritical regime? And, in the affirmative case, does a supercritical regime exist, where the bedform would pass over the obstacle?

In this letter, we investigate experimentally the behavior of subaqueous barchans that interact with dune-size obstacles. For that, we carried out experiments in a water channel where a large 3D obstacle was fixed initially downstream of a barchan dune, which was filmed as it reached the obstacle and interacted with it. In our experiments, we varied the obstacle shape and size, the flow velocity, and the grains' properties. We show that subaqueous barchans can be blocked, bypass, or pass over dune-size obstacles, with some intermediate/transient situations depending on the varied parameters. In the bypass cases, the original barchan can split into two or more bedforms, resulting in a significant redistribution of sand in space. Finally, we propose a classification map in which the barchan behavior depends basically on two dimensionless parameters. Our results represent a step toward understanding how barchans behave in the presence of large obstacles, helping us to predict the outcomes of dunes interacting with objects in other environments, and design structures such as retaining walls and bridge pillars that are safer and do not disturb sandy terrains.

\section{Materials and Methods}

The experimental device consisted basically of a water reservoir, two centrifugal pumps, a flow straightener, a 5-m-long closed-conduit channel, a settling tank, and a return line. The channel had a rectangular cross section (width = 160 mm and height 2$\delta$ = 50 mm) and was made of transparent material, and its last 2 m consisted of the 1-m-long test section followed by a 1-m-long section discharging in the settling tank. This assured that a fully-developed water flow arrived at the test section, which was approximately 40 hydraulic diameters (40 $\times$ 3.05$\delta$) from the channel entrance. With a given obstacle fixed on the bottom wall of the test section (centered in the spanwise direction) and the channel previously filled with water, controlled grains were poured inside, forming conical pile upstream the obstacle. By turning on the centrifugal pumps, a pressure-driven turbulent flow was imposed in the channel, deforming the conical pile into a barchan dune that afterward interacted with the obstacle. A camera placed above the channel acquired images of the bedforms as they interacted with the obstacles, and their morphodynamics was obtained later by processing the images. A layout of the experimental setup, photographs of its parts, a photograph of the test section, and microscopy images of the used grains are shown in Figures S1 to S6 in the Supporting Information.

We used tap water at temperatures within 21 and 25 $^o$C and three populations of solid particles: glass spheres (density $\rho_s$ = 2500 kg/m$^3$) with diameters within $0.40$ mm $\leq\,d\,\leq$ $0.60$ mm, glass spheres with 0.15 mm $\leq\,d\,\leq$ 0.25 mm, and zirconium spheres (density $\rho_s$ = 4100 kg/m$^3$) with diameters within $0.40$ mm $\leq\,d\,\leq$ $0.60$ mm (the latter only for tests with cylindrical obstacles). The cross-sectional mean velocities of water $U$ were within 0.226 m/s and 0.312 m/s, corresponding to Reynolds numbers based on the channel height \cite{Panton}, Re = $\rho U 2\delta /\mu$, within 1.13 $\times$ 10$^4$ and 1.55 $\times$ 10$^4$, respectively, and to Stokes numbers \cite{Andreotti_6} $St_t \,=\, U d \rho_s / (18\mu)$ within 6.3 and 35.5, where $\rho$ is the density and $\mu$ the dynamic viscosity of the fluid. The shear velocities on the channel walls $u_*$ (base flow) were computed from velocity profiles measured in previous works with a two-dimensional particle image velocimetry device \cite{Franklin_9, Cunez2, Alvarez3}, and were found to follow the Blasius correlation \cite{Schlichting_1}, from which we found 0.0133 m/s $\leq$ $u_*$ $\leq$ 0.0168 m/s. This corresponds to Reynolds numbers at the grain scale, Re$_*$ = $\rho u_* d / \mu$, within 3 and 8 and to Shields numbers, $\theta$ = $(\rho u_*^2)/((\rho_s - \rho )gd)$, within 0.060 and 0.086.

The initial mass $m$ of dunes varied within 0.5 and 40.0 g, corresponding to conical piles with basal diameter within 17 and 75 mm. We call in the following $W$ the width of a barchan dune, defined as its spanwise dimension. For the obstacles, we used different geometries (cylinders, rings, blocks and spheres) and sizes. Whenever we mention their height $H_{obs}$ and width $W_{obs}$, we refer to dimensions measured considering a frontal view (taken in the direction of the fluid flow). Photographs of the used objects and a complete list of their sizes, as well as a list of the tested conditions, are available in Figure S7, Table S1, and Figure S16, respectively, in the Supporting Information.

\section{Results}

We basically observed four behaviors when a subaqueous barchan reaches a dune-size obstacle. It either: (i) bypasses the obstacle, without touching its surface; (ii) passes over the obstacle; (iii) behaves in an intermediate situation between passing over and bypassing the obstacle; or (iv) is greatly deformed, with a large portion of grains remaining trapped in the upstream part of the obstacle. We call these behaviors \textit{bypass}, \textit{pass over}, \textit{transient} and \textit{trapping}, respectively, and they do not appear necessarily to all obstacle shapes. Besides the obstacle shape, we observed that the behaviors depend also on the relative size between the dune and the obstacle, the water flow conditions, and the transport conditions (how easily particles follow the fluid). We inquire into the relations between these parameters next.

\begin{figure}[ht]
	\begin{center}
		\includegraphics[width=1\linewidth]{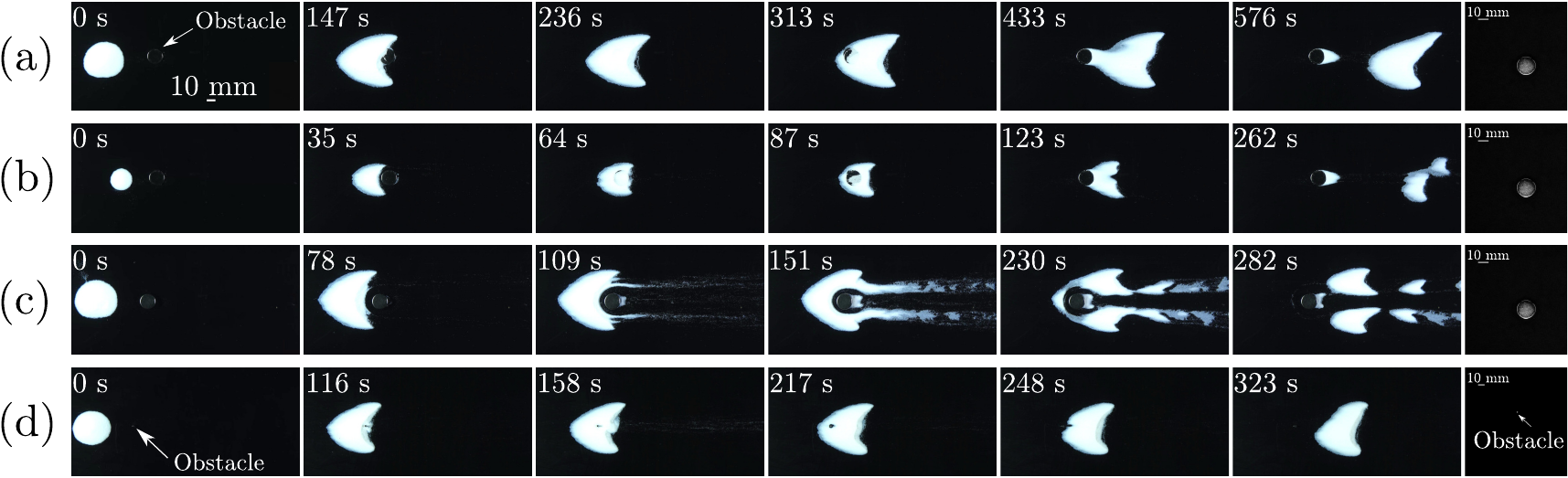}\\
	\end{center}
	\caption{Snapshots of a barchan dune interacting with a cylindrical object, for different size ratios. In the snapshots, the water flow is from left to right, and the corresponding times are shown in each frame. (a) $H_{obst}/W$ = 0.05, for which the dune passes over the obstacle and continues as a single barchan (test 8); (b) $H_{obst}/W$ = 0.08, for which the dune behaves in an intermediate situation between passing over and bypassing the obstacle (test 9); (c) $H_{obst}/W$ = 0.19, for which the dune bypasses the obstacle (test 23); (d) $H_{obst}/W$ = 0.21 (thin cylinder), for which the dune passes over the obstacle and continues as a single barchan (test 47). The obstacle appears as a bright metallic object in each frame. On the right of each sequence, there is a photograph showing a top view of the used obstacle. Movies of each of these sequences and a table listing the test conditions are available in Movies S1 to S4 and Figure S16 in the Supporting Information.}
	\label{fig:snapshots1}
\end{figure}

We begin with the cylinders, which we investigated first. For this form, we observed only the pass over, transient and bypass behaviors, shown in Figure \ref{fig:snapshots1}, which indicates a dependence on both the relative height and diameter of cylinders with respect to the barchan dimensions (snapshots for other cases are available in the supporting information). In general, cylinders with height and diameter comparable to those of barchans produce the bypass behavior, in which the dune circumvents the obstacle with its grains avoiding the obstacle surface. As a result, the barchan is divided into smaller barchans by the end of the dune-obstacle interaction (four in Figure \ref{fig:snapshots1}c). This grain-wall shunning indicates that grains are closely following the water flow, so that local Stokes numbers (how close the behavior of solid and fluid particles are) shall be relatively low. On the other hand, when the cylinder is very thin (small diameter, Figure \ref{fig:snapshots1}d) or low (Figure \ref{fig:snapshots1}a), the dune passes over the obstacle without much disturbance other than leaving part of its grains in the recirculation region just downstream the obstacle. For intermediate sizes, the barchan dune behaves in a transient manner (Figure \ref{fig:snapshots1}b).

\begin{figure}[ht]
	\begin{center}
		\includegraphics[width=1\linewidth]{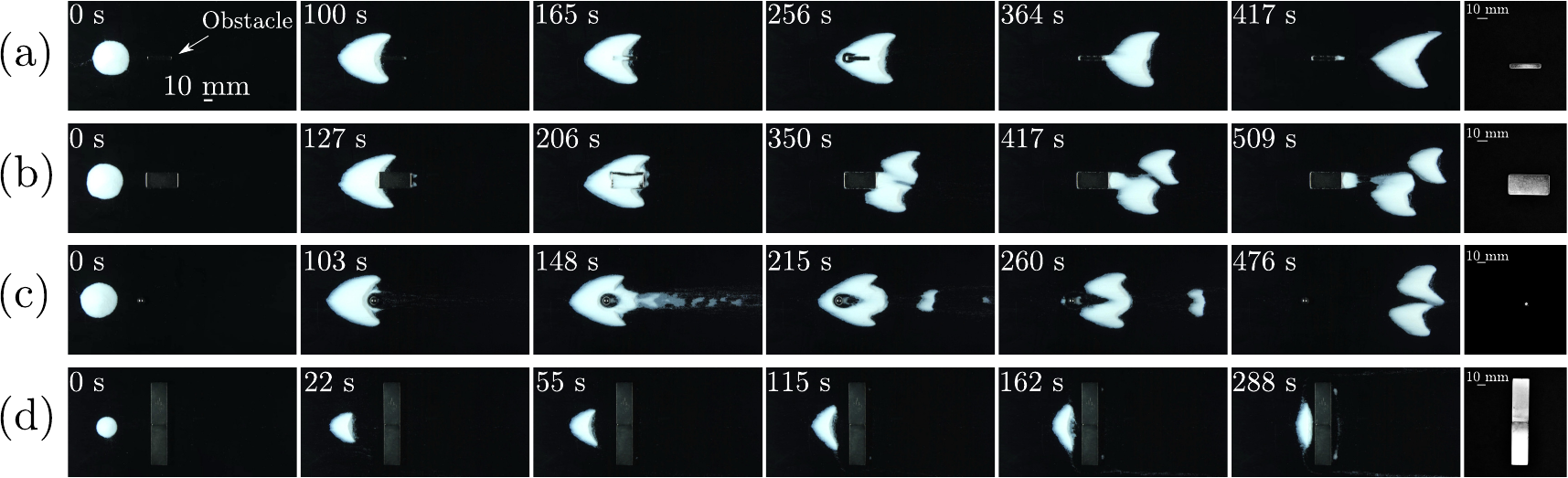}\\
	\end{center}
	\caption{Snapshots of a barchan dune interacting with objects of other shapes. In the snapshots, the water flow is from left to right, and the corresponding times are shown in each frame. (a) Block with $H_{obst}/W$ = 0.12 and $H_{obst}/W_{obst}$ = 1, for which the dune passes over the obstacle and continues as a single barchan (test 41); (b) block with $H_{obst}/W$ = 0.12 and $H_{obst}/W_{obst}$ = 0.28, for which the dune behaves in an intermediate situation between passing over and bypassing the obstacle (test 45); (c) sphere with $H_{obst}/W$ = 0.23, for which the dune bypasses the obstacle (test 33); (d) block with $H_{obst}/W$ = 0.40 and $H_{obst}/W_{obst}$ = 0.10, for which the dune is trapped and spreads in the spanwise direction (test 55). The obstacle appears as a bright metallic object in each frame. On the right of each sequence, there is a photograph showing a top view of the used obstacle. Movies of each of these sequences and a table listing the test conditions are available in Movies S5 to S8 and Figure S16 in the Supporting Information.}
	\label{fig:snapshots2}
\end{figure}

We also carried out experiments with other geometries, namely blocks with different aspect ratios, rings, and spheres (images of all used obstacles are available in Figure S7 in the Supporting Information). Figure \ref{fig:snapshots2} shows snapshots of the barchan dune interacting with blocks and a sphere, from which we observe that trapping  occurs for the wide block (Figure \ref{fig:snapshots2}d). The reason for blocking considerable quantities of grains just upstream the obstacle is the large flow disturbance in the spanwise direction. In its turn, the flow disturbance is caused by the comparable heights of object and dune, and large aspect ratio (largest dimension in the spanwise direction) of the obstable. This large disturbance spreads the grains laterally in the upstream region, with many of them remaining trapped, while others circunvent the obstacle and are either entrained further downstream or kept in the recirculation region just downstream the obstacle (two small spots linked by a thin stripe in Figure \ref{fig:snapshots2}d). For the other obstacles shown in Figure \ref{fig:snapshots2}, the general behavior is similar to those found for cylinders, and we have not observed changes of the type of interaction by varying the streamwise elongation of obstacles.

\begin{figure}[ht]
	\begin{center}
		\includegraphics[width=.99\linewidth]{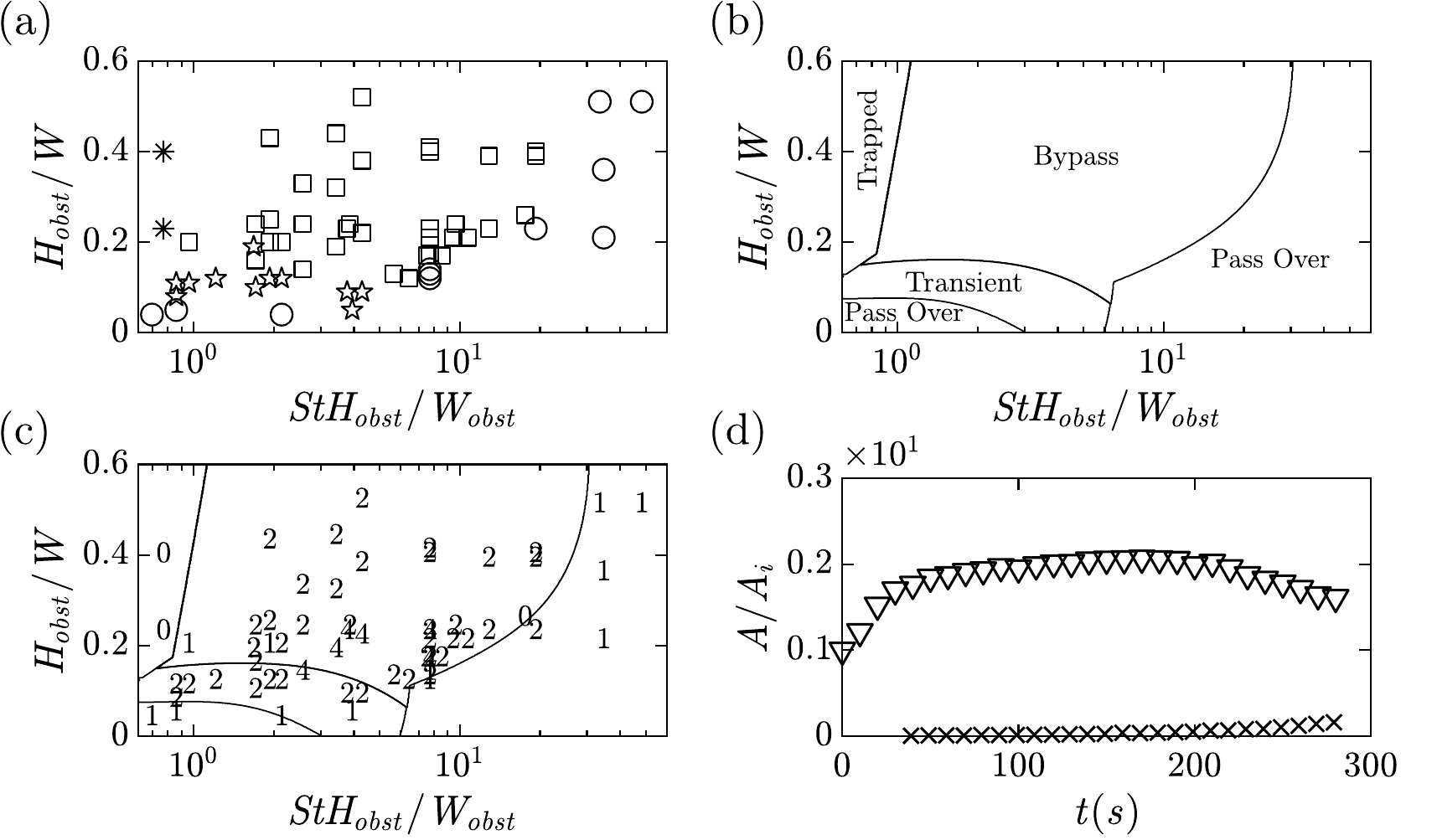}\\
	\end{center}
	\caption{(a) Different behaviors observed in the diagram of size ratio vs. Stokes effect, i.e., $H_{obst}/W$ vs. $St H_{obst} / W_{obst}$. Squares, pentagrams, circles and asterisks correspond to bypass, transient, pass over and trapping patterns. (b) Map showing regions where the different behaviors are expected (based on machine learning applied to the map of panel (a)). (c) Number of resulting bedforms observed in the $H_{obst}/W$ vs. $St H_{obst} / W_{obst}$ diagram (the continuous lines are those of Figure (c)). (d) Projected area of the resulting bedforms normalized by that of the initial pile, for the trapped case of Figure \ref{fig:snapshots2}d (test 55). Triangles correspond to the upstream bedform and crosses to the downstream ones (the two spots plus the transverse stripe).}
	\label{fig:map}
\end{figure}

While the size ratio between obstacle and dune is an important parameter affecting the type of interaction, the grain-wall shunning taking place in bypass cases strongly indicates that the transport of grains by the fluid must be taken into account. This is in agreement with the findings reported by \citeA{Bacik} in the 2D case (although in their case the bypass does not exist and the passing over results only from grains' displacements in the vertical direction). Therefore, in order to rationalize the problem, we inquired into dimensionless parameters regrouping the observed behaviors. We ended up with two dimensionless numbers: the ratio between the obstacle height and the barchan width, $H_{obst}/W$, and the Stokes number multiplied the obstacle height and divided by its width, $St H_{obst} / W_{obst}$. While the first is clearly a ratio between heights \cite<the barchan height being around 10\% of its width, >{Franklin_8}, the second is a measure of how close the grains follow the water flow (closer for smaller values of $St$) and how much the flow is disturbed in the spanwise direction (greater for smaller values of $H_{obst} / W_{obst}$).

Figure \ref{fig:map}a presents a map of the outputs of our experiments plotted in the $H_{obst}/W$ vs. $St H_{obst} / W_{obst}$ space, where the squares, pentagrams, circles and asterisks correspond to the bypass, transient, pass over and trapping behaviors, respectively. The behaviors seem reasonably well organized in the map, where we can observe that:

\begin{itemize}
	
	\item bypass cases occur for high $H_{obst}/W$ (above 0.1) and moderate $St H_{obst} / W_{obst}$ (within 1 and 20) values;
	
	\item pass over cases occur for low to moderate $H_{obst}/W$ and moderate to high $St H_{obst} / W_{obst}$ values ($H_{obst}/W$ $<$ 0.2 for 1 $<$ $St H_{obst} / W_{obst}$ $<$ 10, and 0.2 $\leq$ $H_{obst}/W$ $\leq$ 0.4 for $St H_{obst} / W_{obst}$ $>$ 10);
			
	\item transient cases occur for moderate values of both $H_{obst}/W$ and $St H_{obst} / W_{obst}$, i.e., in the vicinity of the bypass and pass over cases ($H_{obst}/W$ within 0.05 and 0.2, and $St H_{obst} / W_{obst}$ within 1 and 5);
	
	\item and trapping cases occur for high $H_{obst}/W$  (above 0.2) and very low $St H_{obst} / W_{obst}$ (below 1) values;
		
\end{itemize}

Given the organization of patterns in the $H_{obst}/W$ vs. $St H_{obst} / W_{obst}$ space, we inquired into the lines separating the different regions of Figure \ref{fig:map}a. For that, we applied a machine learning algorithm \cite<the Support Vector Machine method,>{Mammone} to Figure \ref{fig:map}a, from which we obtained the separation lines shown in Figure \ref{fig:map}b  (see the Supporting Information for a brief description of the machine learning method Support Vector Machine). Figure \ref{fig:map}b consists, therefore, in a tentative map for classifying the behavior of subaqueous barchans interacting with dune-scale obstacles. We also investigated the number of dunes resulting from the interaction between the barchan and the obstacle. From the snapshots of Figures \ref{fig:snapshots1} and \ref{fig:snapshots2}, we observe that sometimes interactions give rise to four dunes (Fig. \ref{fig:snapshots1}c, for example) or zero (Fig. \ref{fig:snapshots2}d, for example). We identified the number of dunes by the end of each test, and plotted that number directly in the $H_{obst}/W$ vs. $St H_{obst} / W_{obst}$ space, shown in Figure \ref{fig:map}c (which also shows the separation lines). Here again, data is well organized, with roughly four or two dunes for the bypass case, two for the transient case, one for the pass over case, and zero for the trapped case. Maps similar to those of Figures \ref{fig:map}a-\ref{fig:map}c can be obtained by using a modified Shields number $\theta H_{obst} / W_{obst}$ instead of $St H_{obst} / W_{obst}$. However, because the flow is the main responsible for the bypass pattern, we believe that the $St H_{obst} / W_{obst}$ is the proper dimensionless number. Maps using $\theta H_{obst} / W_{obst}$ are available in Figures S11-S12 and S14-S15 in the Supporting Information. 

Finally, we investigated the spreading of grains in the transverse direction (upstream the obstacle) and the entrapment of part of them in the recirculation region downstream the obstacle that occur in the trapping case. For that, we computed the surface area of bedforms and clusters of grains projected in the horizontal plane, and normalized them by the projected area of the initial pile. One example is shown in Figure \ref{fig:map}d, plotting the time evolution of areas of both the upstream bedform (triangles) and downstream clusters (crosses). We observe that the area of the upstream bedform increases during its lateral elongation, reaching approximately the double of the value of the initial area, and afterward it decreases slowly as some grains are entrained downstream (circumventing the obstacle). As the area of the upstream bedform decreases, those of the two spots linked by the transverse stripe increase (downstream the obstacle), but not at the same proportion since some of the grains are entrained further downstream. 

We note that our results are only valid above the threshold for granular motion (otherwise there will be no interaction between dunes and obstacles) and concern only subaqueous bedforms, even though the map of Figure \ref{fig:map}b uses the modified Stokes number. For the subaqueous case, the density ratio between sand and fluid is of the order of unity, so that sand particles are entrained directly by the water flow by rolling, sliding or effectuating small jumps (of the order few grain diameters). The situation is different when the fluid is a gas. In the aeolian case, for example, the density ratio is of the order of 10$^3$, and sand particles effectuate large jumps (much larger that the grain diameter) due to their larger inertia. The grains forming aeolian barchans are, thus, expected to collide with the obstacle even when large flow disturbances are present. Although this is in agreement with the trend shown on the right part of the map, which indicates that high-inertia particles would collide and pass over obstacles, care must be taken when applying the maps of Figures \ref{fig:map}b-\ref{fig:map}c to aeolian or Martian dunes.

\section{Conclusions}

In summary, we found that subaqueous barchans can be blocked, bypass or pass over dune-size obstacles (with some intermediate situations), that blocked dunes spread in the spanwise direction and remain trapped just upstream the obstacle, and that bypassing barchans split in two or more bedforms, implying a significant redistribution of sand in space. Based on the experiments, we propose a classification map in which the barchan behavior depends on two dimensionless parameters: an obstacle-dune size ratio and a modified Stokes number. Our results shed light on how barchans behave in the presence of large obstacles, helping us to: (i) better understand the flow effects on barchan-barchan interactions \cite{Assis}; (ii) predict the outcomes of dunes interacting with objects in other environments; and (iii) design structures such as retaining walls and bridge pillars that are safer and do not disturb sandy terrains.

\section*{Open Research}
\begin{sloppypar}
	Data (digital images) supporting this work were generated by ourselves and are available in Mendeley Data \cite{Supplemental2} under the CC-BY-4.0 license. The numerical scripts used to process the images are also available in Mendeley Data \cite{Supplemental2} under the CC-BY-4.0 license.
\end{sloppypar}

\acknowledgments
\begin{sloppypar}
The authors are grateful to FAPESP (Grant Nos. 2018/14981-7, 2019/10239-7 and 2022/01758-3) and to CNPq (Grant No. 405512/2022-8) for the financial support provided.
\end{sloppypar}

\nocite{cover_sup, kowalczyk_sup, Pedregosa_sup, rhys_sup}
\bibliography{references}

\begin{thebibliography}{}

\bibitem [\protect \citeauthoryear {%
Alvarez%
, C\'u\~nez%
\BCBL {}\ \BBA {} Franklin%
}{%
Alvarez%
\ \protect \BOthers {.}}{%
{\protect \APACyear {2021}}%
}]{%
Alvarez6}
\APACinsertmetastar {%
Alvarez6}%
\begin{APACrefauthors}%
Alvarez, C\BPBI A.%
, C\'u\~nez, F\BPBI D.%
\BCBL {}\ \BBA {} Franklin, E\BPBI M.%
\end{APACrefauthors}%
\unskip\
\newblock
\APACrefYearMonthDay{2021}{}{}.
\newblock
{\BBOQ}\APACrefatitle {Growth of barchan dunes of bidispersed granular
  mixtures} {Growth of barchan dunes of bidispersed granular mixtures}.{\BBCQ}
\newblock
\APACjournalVolNumPages{Phys. Fluids}{33}{5}{051705}.
\PrintBackRefs{\CurrentBib}

\bibitem [\protect \citeauthoryear {%
Alvarez%
\ \BBA {} Franklin%
}{%
Alvarez%
\ \BBA {} Franklin%
}{%
{\protect \APACyear {2018}}%
}]{%
Alvarez3}
\APACinsertmetastar {%
Alvarez3}%
\begin{APACrefauthors}%
Alvarez, C\BPBI A.%
\BCBT {}\ \BBA {} Franklin, E\BPBI M.%
\end{APACrefauthors}%
\unskip\
\newblock
\APACrefYearMonthDay{2018}{Oct}{}.
\newblock
{\BBOQ}\APACrefatitle {Role of Transverse Displacements in the Formation of
  Subaqueous Barchan Dunes} {Role of transverse displacements in the formation
  of subaqueous barchan dunes}.{\BBCQ}
\newblock
\APACjournalVolNumPages{Phys. Rev. Lett.}{121}{}{164503}.
\newblock
\begin{APACrefURL}
  \url{https://link.aps.org/doi/10.1103/PhysRevLett.121.164503}
  \end{APACrefURL}
\newblock
\begin{APACrefDOI} \doi{10.1103/PhysRevLett.121.164503} \end{APACrefDOI}
\PrintBackRefs{\CurrentBib}

\bibitem [\protect \citeauthoryear {%
Amaral%
, dos Santos%
, Ribeiro%
\BCBL {}\ \BBA {} Barreto%
}{%
Amaral%
\ \protect \BOthers {.}}{%
{\protect \APACyear {2019}}%
}]{%
Amaral}
\APACinsertmetastar {%
Amaral}%
\begin{APACrefauthors}%
Amaral, Y\BPBI T.%
, dos Santos, E\BPBI M.%
, Ribeiro, M\BPBI C.%
\BCBL {}\ \BBA {} Barreto, L.%
\end{APACrefauthors}%
\unskip\
\newblock
\APACrefYearMonthDay{2019}{}{}.
\newblock
{\BBOQ}\APACrefatitle {Landscape structural analysis of the Lençóis
  Maranhenses national park: implications for conservation} {Landscape
  structural analysis of the lençóis maranhenses national park: implications
  for conservation}.{\BBCQ}
\newblock
\APACjournalVolNumPages{J. Nat. Conserv.}{51}{}{125725}.
\PrintBackRefs{\CurrentBib}

\bibitem [\protect \citeauthoryear {%
Andreotti%
, Forterre%
\BCBL {}\ \BBA {} Pouliquen%
}{%
Andreotti%
\ \protect \BOthers {.}}{%
{\protect \APACyear {2013}}%
}]{%
Andreotti_6}
\APACinsertmetastar {%
Andreotti_6}%
\begin{APACrefauthors}%
Andreotti, B.%
, Forterre, Y.%
\BCBL {}\ \BBA {} Pouliquen, O.%
\end{APACrefauthors}%
\unskip\
\newblock
\APACrefYear{2013}.
\newblock
\APACrefbtitle {Granular Flow: {B}etween Fluid and Solid} {Granular flow:
  {B}etween fluid and solid}.
\newblock
\APACaddressPublisher{}{Cambridge University Press}.
\PrintBackRefs{\CurrentBib}

\bibitem [\protect \citeauthoryear {%
Assis%
, Borges%
\BCBL {}\ \BBA {} Franklin%
}{%
Assis%
\ \protect \BOthers {.}}{%
{\protect \APACyear {2023}}%
}]{%
Supplemental2}
\APACinsertmetastar {%
Supplemental2}%
\begin{APACrefauthors}%
Assis, W\BPBI R.%
, Borges, D\BPBI S.%
\BCBL {}\ \BBA {} Franklin, E\BPBI M.%
\end{APACrefauthors}%
\unskip\
\newblock
\APACrefYearMonthDay{2023}{}{}.
\newblock
{\BBOQ}\APACrefatitle {Dataset for ``{B}archan dunes cruising dune-size
  obstacles'' [{D}ataset][{S}oftware]} {Dataset for ``{B}archan dunes cruising
  dune-size obstacles'' [{D}ataset][{S}oftware]}.{\BBCQ}
\newblock
\APACjournalVolNumPages{Mendeley Data,
  http://dx.doi.org/10.17632/snffc3wvfp.1}{}{}{}.
\newblock
\begin{APACrefDOI} \doi{10.17632/snffc3wvfp.1} \end{APACrefDOI}
\PrintBackRefs{\CurrentBib}

\bibitem [\protect \citeauthoryear {%
Assis%
, Cúñez%
\BCBL {}\ \BBA {} Franklin%
}{%
Assis%
\ \protect \BOthers {.}}{%
{\protect \APACyear {2022}}%
}]{%
Assis3}
\APACinsertmetastar {%
Assis3}%
\begin{APACrefauthors}%
Assis, W\BPBI R.%
, Cúñez, F\BPBI D.%
\BCBL {}\ \BBA {} Franklin, E\BPBI M.%
\end{APACrefauthors}%
\unskip\
\newblock
\APACrefYearMonthDay{2022}{}{}.
\newblock
{\BBOQ}\APACrefatitle {Revealing the Intricate Dune-Dune Interactions of
  Bidisperse Barchans} {Revealing the intricate dune-dune interactions of
  bidisperse barchans}.{\BBCQ}
\newblock
\APACjournalVolNumPages{J. Geophys. Res.: Earth Surf.}{127}{5}{e2021JF006588}.
\newblock
\begin{APACrefURL}
  \url{https://agupubs.onlinelibrary.wiley.com/doi/abs/10.1029/2021JF006588}
  \end{APACrefURL}
\newblock
\begin{APACrefDOI} \doi{https://doi.org/10.1029/2021JF006588} \end{APACrefDOI}
\PrintBackRefs{\CurrentBib}

\bibitem [\protect \citeauthoryear {%
Assis%
\ \BBA {} Franklin%
}{%
Assis%
\ \BBA {} Franklin%
}{%
{\protect \APACyear {2020}}%
}]{%
Assis}
\APACinsertmetastar {%
Assis}%
\begin{APACrefauthors}%
Assis, W\BPBI R.%
\BCBT {}\ \BBA {} Franklin, E\BPBI M.%
\end{APACrefauthors}%
\unskip\
\newblock
\APACrefYearMonthDay{2020}{}{}.
\newblock
{\BBOQ}\APACrefatitle {A Comprehensive Picture for Binary Interactions of
  Subaqueous Barchans} {A comprehensive picture for binary interactions of
  subaqueous barchans}.{\BBCQ}
\newblock
\APACjournalVolNumPages{Geophys. Res. Lett.}{47}{18}{e2020GL089464}.
\PrintBackRefs{\CurrentBib}

\bibitem [\protect \citeauthoryear {%
Bacik%
, Canizares%
, Caulfield%
, Williams%
\BCBL {}\ \BBA {} Vriend%
}{%
Bacik%
\ \protect \BOthers {.}}{%
{\protect \APACyear {2021}}%
}]{%
Bacik}
\APACinsertmetastar {%
Bacik}%
\begin{APACrefauthors}%
Bacik, K\BPBI A.%
, Canizares, P.%
, Caulfield, C\BHBI c\BPBI P.%
, Williams, M\BPBI J.%
\BCBL {}\ \BBA {} Vriend, N\BPBI M.%
\end{APACrefauthors}%
\unskip\
\newblock
\APACrefYearMonthDay{2021}{Oct}{}.
\newblock
{\BBOQ}\APACrefatitle {Dynamics of migrating sand dunes interacting with
  obstacles} {Dynamics of migrating sand dunes interacting with
  obstacles}.{\BBCQ}
\newblock
\APACjournalVolNumPages{Phys. Rev. Fluids}{6}{}{104308}.
\newblock
\begin{APACrefURL}
  \url{https://link.aps.org/doi/10.1103/PhysRevFluids.6.104308}
  \end{APACrefURL}
\newblock
\begin{APACrefDOI} \doi{10.1103/PhysRevFluids.6.104308} \end{APACrefDOI}
\PrintBackRefs{\CurrentBib}

\bibitem [\protect \citeauthoryear {%
Bagnold%
}{%
Bagnold%
}{%
{\protect \APACyear {1941}}%
}]{%
Bagnold_1}
\APACinsertmetastar {%
Bagnold_1}%
\begin{APACrefauthors}%
Bagnold, R\BPBI A.%
\end{APACrefauthors}%
\unskip\
\newblock
\APACrefYear{1941}.
\newblock
\APACrefbtitle {The Physics of Blown Sand and Desert Dunes} {The physics of
  blown sand and desert dunes}.
\newblock
\APACaddressPublisher{London}{Chapman and Hall}.
\PrintBackRefs{\CurrentBib}

\bibitem [\protect \citeauthoryear {%
Bourke%
}{%
Bourke%
}{%
{\protect \APACyear {2010}}%
}]{%
Bourke}
\APACinsertmetastar {%
Bourke}%
\begin{APACrefauthors}%
Bourke, M\BPBI C.%
\end{APACrefauthors}%
\unskip\
\newblock
\APACrefYearMonthDay{2010}{}{}.
\newblock
{\BBOQ}\APACrefatitle {Barchan dune asymmetry: Observations from {M}ars and
  {E}arth} {Barchan dune asymmetry: Observations from {M}ars and
  {E}arth}.{\BBCQ}
\newblock
\APACjournalVolNumPages{Icarus}{205}{1}{183 - 197}.
\PrintBackRefs{\CurrentBib}

\bibitem [\protect \citeauthoryear {%
Breed%
, Grolier%
\BCBL {}\ \BBA {} McCauley%
}{%
Breed%
\ \protect \BOthers {.}}{%
{\protect \APACyear {1979}}%
}]{%
Breed}
\APACinsertmetastar {%
Breed}%
\begin{APACrefauthors}%
Breed, C\BPBI S.%
, Grolier, M\BPBI J.%
\BCBL {}\ \BBA {} McCauley, J\BPBI F.%
\end{APACrefauthors}%
\unskip\
\newblock
\APACrefYearMonthDay{1979}{}{}.
\newblock
{\BBOQ}\APACrefatitle {Morphology and distribution of common ‘sand’ dunes
  on {M}ars: {C}omparison with the {E}arth} {Morphology and distribution of
  common ‘sand’ dunes on {M}ars: {C}omparison with the {E}arth}.{\BBCQ}
\newblock
\APACjournalVolNumPages{J. Geophys. Res.: Sol. Earth}{84}{B14}{8183-8204}.
\PrintBackRefs{\CurrentBib}

\bibitem [\protect \citeauthoryear {%
Claudin%
\ \BBA {} Andreotti%
}{%
Claudin%
\ \BBA {} Andreotti%
}{%
{\protect \APACyear {2006}}%
}]{%
Claudin_Andreotti}
\APACinsertmetastar {%
Claudin_Andreotti}%
\begin{APACrefauthors}%
Claudin, P.%
\BCBT {}\ \BBA {} Andreotti, B.%
\end{APACrefauthors}%
\unskip\
\newblock
\APACrefYearMonthDay{2006}{}{}.
\newblock
{\BBOQ}\APACrefatitle {A scaling law for aeolian dunes on {M}ars, {V}enus,
  {E}arth, and for subaqueous ripples} {A scaling law for aeolian dunes on
  {M}ars, {V}enus, {E}arth, and for subaqueous ripples}.{\BBCQ}
\newblock
\APACjournalVolNumPages{Earth Plan. Sci. Lett.}{252}{}{20-44}.
\PrintBackRefs{\CurrentBib}

\bibitem [\protect \citeauthoryear {%
{Courrech du Pont}%
}{%
{Courrech du Pont}%
}{%
{\protect \APACyear {2015}}%
}]{%
Courrech}
\APACinsertmetastar {%
Courrech}%
\begin{APACrefauthors}%
{Courrech du Pont}, S.%
\end{APACrefauthors}%
\unskip\
\newblock
\APACrefYearMonthDay{2015}{}{}.
\newblock
{\BBOQ}\APACrefatitle {Dune morphodynamics} {Dune morphodynamics}.{\BBCQ}
\newblock
\APACjournalVolNumPages{C. R. Phys.}{16}{1}{118 - 138}.
\PrintBackRefs{\CurrentBib}

\bibitem [\protect \citeauthoryear {%
Cover%
}{%
Cover%
}{%
{\protect \APACyear {1965}}%
}]{%
cover_sup}
\APACinsertmetastar {%
cover_sup}%
\begin{APACrefauthors}%
Cover, T\BPBI M.%
\end{APACrefauthors}%
\unskip\
\newblock
\APACrefYearMonthDay{1965}{}{}.
\newblock
{\BBOQ}\APACrefatitle {Geometrical and Statistical Properties of Systems of
  Linear Inequalities with Applications in Pattern Recognition} {Geometrical
  and statistical properties of systems of linear inequalities with
  applications in pattern recognition}.{\BBCQ}
\newblock
\APACjournalVolNumPages{IEEE Transactions on Electronic
  Computers}{EC-14}{3}{326-334}.
\newblock
\begin{APACrefDOI} \doi{10.1109/PGEC.1965.264137} \end{APACrefDOI}
\PrintBackRefs{\CurrentBib}

\bibitem [\protect \citeauthoryear {%
C\'u\~nez%
, Oliveira%
\BCBL {}\ \BBA {} Franklin%
}{%
C\'u\~nez%
\ \protect \BOthers {.}}{%
{\protect \APACyear {2018}}%
}]{%
Cunez2}
\APACinsertmetastar {%
Cunez2}%
\begin{APACrefauthors}%
C\'u\~nez, F\BPBI D.%
, Oliveira, G\BPBI V\BPBI G.%
\BCBL {}\ \BBA {} Franklin, E\BPBI M.%
\end{APACrefauthors}%
\unskip\
\newblock
\APACrefYearMonthDay{2018}{}{}.
\newblock
{\BBOQ}\APACrefatitle {Turbulent channel flow perturbed by triangular ripples}
  {Turbulent channel flow perturbed by triangular ripples}.{\BBCQ}
\newblock
\APACjournalVolNumPages{J. Braz. Soc. Mech. Sci. Eng.}{40}{138}{}.
\PrintBackRefs{\CurrentBib}

\bibitem [\protect \citeauthoryear {%
Elbelrhiti%
, Claudin%
\BCBL {}\ \BBA {} Andreotti%
}{%
Elbelrhiti%
\ \protect \BOthers {.}}{%
{\protect \APACyear {2005}}%
}]{%
Elbelrhiti}
\APACinsertmetastar {%
Elbelrhiti}%
\begin{APACrefauthors}%
Elbelrhiti, H.%
, Claudin, P.%
\BCBL {}\ \BBA {} Andreotti, B.%
\end{APACrefauthors}%
\unskip\
\newblock
\APACrefYearMonthDay{2005}{}{}.
\newblock
{\BBOQ}\APACrefatitle {Field evidence for surface-wave-induced instability of
  sand dunes} {Field evidence for surface-wave-induced instability of sand
  dunes}.{\BBCQ}
\newblock
\APACjournalVolNumPages{Nature}{437}{04058}{}.
\PrintBackRefs{\CurrentBib}

\bibitem [\protect \citeauthoryear {%
Finkel%
}{%
Finkel%
}{%
{\protect \APACyear {1959}}%
}]{%
Finkel}
\APACinsertmetastar {%
Finkel}%
\begin{APACrefauthors}%
Finkel, H\BPBI J.%
\end{APACrefauthors}%
\unskip\
\newblock
\APACrefYearMonthDay{1959}{}{}.
\newblock
{\BBOQ}\APACrefatitle {The barchans of southern Peru} {The barchans of southern
  peru}.{\BBCQ}
\newblock
\APACjournalVolNumPages{J. Geol.}{67}{6}{614-647}.
\PrintBackRefs{\CurrentBib}

\bibitem [\protect \citeauthoryear {%
Franklin%
\ \BBA {} Charru%
}{%
Franklin%
\ \BBA {} Charru%
}{%
{\protect \APACyear {2011}}%
}]{%
Franklin_8}
\APACinsertmetastar {%
Franklin_8}%
\begin{APACrefauthors}%
Franklin, E\BPBI M.%
\BCBT {}\ \BBA {} Charru, F.%
\end{APACrefauthors}%
\unskip\
\newblock
\APACrefYearMonthDay{2011}{}{}.
\newblock
{\BBOQ}\APACrefatitle {Subaqueous barchan dunes in turbulent shear flow. {P}art
  1. {D}une motion} {Subaqueous barchan dunes in turbulent shear flow. {P}art
  1. {D}une motion}.{\BBCQ}
\newblock
\APACjournalVolNumPages{J. Fluid Mech.}{675}{}{199-222}.
\PrintBackRefs{\CurrentBib}

\bibitem [\protect \citeauthoryear {%
Franklin%
, Figueiredo%
\BCBL {}\ \BBA {} Rosa%
}{%
Franklin%
\ \protect \BOthers {.}}{%
{\protect \APACyear {2014}}%
}]{%
Franklin_9}
\APACinsertmetastar {%
Franklin_9}%
\begin{APACrefauthors}%
Franklin, E\BPBI M.%
, Figueiredo, F\BPBI T.%
\BCBL {}\ \BBA {} Rosa, E\BPBI S.%
\end{APACrefauthors}%
\unskip\
\newblock
\APACrefYearMonthDay{2014}{}{}.
\newblock
{\BBOQ}\APACrefatitle {The feedback effect caused by bed load on a turbulent
  liquid flow} {The feedback effect caused by bed load on a turbulent liquid
  flow}.{\BBCQ}
\newblock
\APACjournalVolNumPages{J. Braz. Soc. Mech. Sci. Eng.}{36}{}{725-736}.
\PrintBackRefs{\CurrentBib}

\bibitem [\protect \citeauthoryear {%
Herrmann%
\ \BBA {} Sauermann%
}{%
Herrmann%
\ \BBA {} Sauermann%
}{%
{\protect \APACyear {2000}}%
}]{%
Herrmann_Sauermann}
\APACinsertmetastar {%
Herrmann_Sauermann}%
\begin{APACrefauthors}%
Herrmann, H\BPBI J.%
\BCBT {}\ \BBA {} Sauermann, G.%
\end{APACrefauthors}%
\unskip\
\newblock
\APACrefYearMonthDay{2000}{}{}.
\newblock
{\BBOQ}\APACrefatitle {The shape of dunes} {The shape of dunes}.{\BBCQ}
\newblock
\APACjournalVolNumPages{Physica A (Amsterdam)}{283}{}{24-30}.
\PrintBackRefs{\CurrentBib}

\bibitem [\protect \citeauthoryear {%
Hersen%
}{%
Hersen%
}{%
{\protect \APACyear {2004}}%
}]{%
Hersen_3}
\APACinsertmetastar {%
Hersen_3}%
\begin{APACrefauthors}%
Hersen, P.%
\end{APACrefauthors}%
\unskip\
\newblock
\APACrefYearMonthDay{2004}{}{}.
\newblock
{\BBOQ}\APACrefatitle {On the crescentic shape of barchan dunes} {On the
  crescentic shape of barchan dunes}.{\BBCQ}
\newblock
\APACjournalVolNumPages{Eur. Phys. J. B}{37}{4}{507--514}.
\PrintBackRefs{\CurrentBib}

\bibitem [\protect \citeauthoryear {%
Hersen%
\ \BBA {} Douady%
}{%
Hersen%
\ \BBA {} Douady%
}{%
{\protect \APACyear {2005}}%
}]{%
Hersen_5}
\APACinsertmetastar {%
Hersen_5}%
\begin{APACrefauthors}%
Hersen, P.%
\BCBT {}\ \BBA {} Douady, S.%
\end{APACrefauthors}%
\unskip\
\newblock
\APACrefYearMonthDay{2005}{}{}.
\newblock
{\BBOQ}\APACrefatitle {Collision of barchan dunes as a mechanism of size
  regulation} {Collision of barchan dunes as a mechanism of size
  regulation}.{\BBCQ}
\newblock
\APACjournalVolNumPages{Geophys. Res. Lett.}{32}{21}{}.
\PrintBackRefs{\CurrentBib}

\bibitem [\protect \citeauthoryear {%
Hersen%
, Douady%
\BCBL {}\ \BBA {} Andreotti%
}{%
Hersen%
\ \protect \BOthers {.}}{%
{\protect \APACyear {2002}}%
}]{%
Hersen_1}
\APACinsertmetastar {%
Hersen_1}%
\begin{APACrefauthors}%
Hersen, P.%
, Douady, S.%
\BCBL {}\ \BBA {} Andreotti, B.%
\end{APACrefauthors}%
\unskip\
\newblock
\APACrefYearMonthDay{2002}{Dec}{}.
\newblock
{\BBOQ}\APACrefatitle {Relevant Length Scale of Barchan Dunes} {Relevant length
  scale of barchan dunes}.{\BBCQ}
\newblock
\APACjournalVolNumPages{Phys. Rev. Lett.}{89}{}{264301}.
\newblock
\begin{APACrefURL} \url{https://link.aps.org/doi/10.1103/PhysRevLett.89.264301}
  \end{APACrefURL}
\newblock
\begin{APACrefDOI} \doi{10.1103/PhysRevLett.89.264301} \end{APACrefDOI}
\PrintBackRefs{\CurrentBib}

\bibitem [\protect \citeauthoryear {%
Kowalczyk%
}{%
Kowalczyk%
}{%
{\protect \APACyear {2017}}%
}]{%
kowalczyk_sup}
\APACinsertmetastar {%
kowalczyk_sup}%
\begin{APACrefauthors}%
Kowalczyk, A.%
\end{APACrefauthors}%
\unskip\
\newblock
\APACrefYear{2017}.
\newblock
\APACrefbtitle {Support vector machines succinctly} {Support vector machines
  succinctly}.
\newblock
\APACaddressPublisher{}{Syncfusion Inc}.
\PrintBackRefs{\CurrentBib}

\bibitem [\protect \citeauthoryear {%
Long%
\ \BBA {} Sharp%
}{%
Long%
\ \BBA {} Sharp%
}{%
{\protect \APACyear {1964}}%
}]{%
Long}
\APACinsertmetastar {%
Long}%
\begin{APACrefauthors}%
Long, J.%
\BCBT {}\ \BBA {} Sharp, R.%
\end{APACrefauthors}%
\unskip\
\newblock
\APACrefYearMonthDay{1964}{}{}.
\newblock
{\BBOQ}\APACrefatitle {Barchan-dune movement in Imperiel Valley, California}
  {Barchan-dune movement in imperiel valley, california}.{\BBCQ}
\newblock
\APACjournalVolNumPages{Bull. Geol. Soc. Am.}{75}{}{149-156}.
\PrintBackRefs{\CurrentBib}

\bibitem [\protect \citeauthoryear {%
Mammone%
, Turchi%
\BCBL {}\ \BBA {} Cristianini%
}{%
Mammone%
\ \protect \BOthers {.}}{%
{\protect \APACyear {2009}}%
}]{%
Mammone}
\APACinsertmetastar {%
Mammone}%
\begin{APACrefauthors}%
Mammone, A.%
, Turchi, M.%
\BCBL {}\ \BBA {} Cristianini, N.%
\end{APACrefauthors}%
\unskip\
\newblock
\APACrefYearMonthDay{2009}{}{}.
\newblock
{\BBOQ}\APACrefatitle {Support vector machines} {Support vector
  machines}.{\BBCQ}
\newblock
\APACjournalVolNumPages{WIREs Computational Statistics}{1}{3}{283-289}.
\PrintBackRefs{\CurrentBib}

\bibitem [\protect \citeauthoryear {%
Panton%
}{%
Panton%
}{%
{\protect \APACyear {2010}}%
}]{%
Panton}
\APACinsertmetastar {%
Panton}%
\begin{APACrefauthors}%
Panton, R\BPBI L.%
\end{APACrefauthors}%
\unskip\
\newblock
\APACrefYear{2010}.
\newblock
\APACrefbtitle {Incompressible Flow} {Incompressible flow}.
\newblock
\APACaddressPublisher{}{John Wiley and Sons}.
\newblock
\begin{APACrefDOI} \doi{10.1002/9781118713075} \end{APACrefDOI}
\PrintBackRefs{\CurrentBib}

\bibitem [\protect \citeauthoryear {%
Parteli%
\ \protect \BOthers {.}}{%
Parteli%
\ \protect \BOthers {.}}{%
{\protect \APACyear {2014}}%
}]{%
Parteli4}
\APACinsertmetastar {%
Parteli4}%
\begin{APACrefauthors}%
Parteli, E\BPBI J\BPBI R.%
, Dur{\'a}n, O.%
, Bourke, M\BPBI C.%
, Tsoar, H.%
, P{\"o}schel, T.%
\BCBL {}\ \BBA {} Herrmann, H.%
\end{APACrefauthors}%
\unskip\
\newblock
\APACrefYearMonthDay{2014}{}{}.
\newblock
{\BBOQ}\APACrefatitle {Origins of barchan dune asymmetry: Insights from
  numerical simulations} {Origins of barchan dune asymmetry: Insights from
  numerical simulations}.{\BBCQ}
\newblock
\APACjournalVolNumPages{Aeol. Res.}{12}{}{121--133}.
\PrintBackRefs{\CurrentBib}

\bibitem [\protect \citeauthoryear {%
Parteli%
\ \BBA {} Herrmann%
}{%
Parteli%
\ \BBA {} Herrmann%
}{%
{\protect \APACyear {2007}}%
}]{%
Parteli2}
\APACinsertmetastar {%
Parteli2}%
\begin{APACrefauthors}%
Parteli, E\BPBI J\BPBI R.%
\BCBT {}\ \BBA {} Herrmann, H\BPBI J.%
\end{APACrefauthors}%
\unskip\
\newblock
\APACrefYearMonthDay{2007}{Oct}{}.
\newblock
{\BBOQ}\APACrefatitle {Dune formation on the present Mars} {Dune formation on
  the present mars}.{\BBCQ}
\newblock
\APACjournalVolNumPages{Phys. Rev. E}{76}{}{041307}.
\newblock
\begin{APACrefURL} \url{https://link.aps.org/doi/10.1103/PhysRevE.76.041307}
  \end{APACrefURL}
\newblock
\begin{APACrefDOI} \doi{10.1103/PhysRevE.76.041307} \end{APACrefDOI}
\PrintBackRefs{\CurrentBib}

\bibitem [\protect \citeauthoryear {%
Pedregosa%
\ \protect \BOthers {.}}{%
Pedregosa%
\ \protect \BOthers {.}}{%
{\protect \APACyear {2011}}%
}]{%
Pedregosa_sup}
\APACinsertmetastar {%
Pedregosa_sup}%
\begin{APACrefauthors}%
Pedregosa, F.%
, Varoquaux, G.%
, Gramfort, A.%
, Michel, V.%
, Thirion, B.%
, Grisel, O.%
\BDBL {}Duchesnay, E.%
\end{APACrefauthors}%
\unskip\
\newblock
\APACrefYearMonthDay{2011}{}{}.
\newblock
{\BBOQ}\APACrefatitle {Scikit-learn: Machine Learning in {P}ython}
  {Scikit-learn: Machine learning in {P}ython}.{\BBCQ}
\newblock
\APACjournalVolNumPages{Journal of Machine Learning
  Research}{12}{}{2825--2830}.
\newblock
\begin{APACrefURL} \url{http://jmlr.org/papers/v12/pedregosa11a.html}
  \end{APACrefURL}
\PrintBackRefs{\CurrentBib}

\bibitem [\protect \citeauthoryear {%
Rhys%
}{%
Rhys%
}{%
{\protect \APACyear {2020}}%
}]{%
rhys_sup}
\APACinsertmetastar {%
rhys_sup}%
\begin{APACrefauthors}%
Rhys, H.%
\end{APACrefauthors}%
\unskip\
\newblock
\APACrefYear{2020}.
\newblock
\APACrefbtitle {Machine Learning with {R}, the tidyverse, and mlr} {Machine
  learning with {R}, the tidyverse, and mlr}.
\newblock
\APACaddressPublisher{}{Manning Publications}.
\PrintBackRefs{\CurrentBib}

\bibitem [\protect \citeauthoryear {%
Roback%
, Runyon%
\BCBL {}\ \BBA {} Avouac%
}{%
Roback%
\ \protect \BOthers {.}}{%
{\protect \APACyear {2020}}%
}]{%
Roback}
\APACinsertmetastar {%
Roback}%
\begin{APACrefauthors}%
Roback, K.%
, Runyon, K.%
\BCBL {}\ \BBA {} Avouac, J.%
\end{APACrefauthors}%
\unskip\
\newblock
\APACrefYearMonthDay{2020}{}{}.
\newblock
{\BBOQ}\APACrefatitle {Craters as sand traps: Dynamics, history, and morphology
  of modern sand transport in an active martian dune field} {Craters as sand
  traps: Dynamics, history, and morphology of modern sand transport in an
  active martian dune field}.{\BBCQ}
\newblock
\APACjournalVolNumPages{Icarus}{342}{}{113642}.
\newblock
\APACrefnote{Current and Recent Landscape Evolution on Mars}
\PrintBackRefs{\CurrentBib}

\bibitem [\protect \citeauthoryear {%
Schlichting%
}{%
Schlichting%
}{%
{\protect \APACyear {2000}}%
}]{%
Schlichting_1}
\APACinsertmetastar {%
Schlichting_1}%
\begin{APACrefauthors}%
Schlichting, H.%
\end{APACrefauthors}%
\unskip\
\newblock
\APACrefYear{2000}.
\newblock
\APACrefbtitle {Boundary-Layer Theory} {Boundary-layer theory}.
\newblock
\APACaddressPublisher{New York}{Springer}.
\PrintBackRefs{\CurrentBib}

\bibitem [\protect \citeauthoryear {%
Urso%
, Chojnacki%
\BCBL {}\ \BBA {} Vaz%
}{%
Urso%
\ \protect \BOthers {.}}{%
{\protect \APACyear {2018}}%
}]{%
Urso}
\APACinsertmetastar {%
Urso}%
\begin{APACrefauthors}%
Urso, A.%
, Chojnacki, M.%
\BCBL {}\ \BBA {} Vaz, D\BPBI A.%
\end{APACrefauthors}%
\unskip\
\newblock
\APACrefYearMonthDay{2018}{}{}.
\newblock
{\BBOQ}\APACrefatitle {Dune-Yardang Interactions in Becquerel Crater, Mars}
  {Dune-yardang interactions in becquerel crater, mars}.{\BBCQ}
\newblock
\APACjournalVolNumPages{J. Geophys. Res.: Planets}{123}{2}{353-368}.
\PrintBackRefs{\CurrentBib}

\bibitem [\protect \citeauthoryear {%
Vermeesch%
}{%
Vermeesch%
}{%
{\protect \APACyear {2011}}%
}]{%
Vermeesch}
\APACinsertmetastar {%
Vermeesch}%
\begin{APACrefauthors}%
Vermeesch, P.%
\end{APACrefauthors}%
\unskip\
\newblock
\APACrefYearMonthDay{2011}{}{}.
\newblock
{\BBOQ}\APACrefatitle {Solitary wave behavior in sand dunes observed from
  space} {Solitary wave behavior in sand dunes observed from space}.{\BBCQ}
\newblock
\APACjournalVolNumPages{Geophys. Res. Lett.}{38}{22}{}.
\PrintBackRefs{\CurrentBib}

\end{thebibliography}

\end{document}